\documentstyle[aps,prb,epsfig]{revtex}
\draft

\begin{document}
%\psdraft

\title{Growth of (110) Diamond using pure Dicarbon}
\author{M. Sternberg$^{1,}$\cite{mailto}, M. Kaukonen$^2$,
    R. M. Nieminen$^2$, Th. Frauenheim$^1$}
\address{$^1$Department of Physics, Theoretical Physics,
    University of Paderborn, D-33098 Paderborn, Germany }
\address{$^2$Laboratory of Physics, Helsinki University of Technology,
    P.O. Box 1100, FIN-02015, Finland }

\maketitle

\begin{abstract}
    We use a density-functional based tight-binding method to study
    diamond growth steps by depositing dicarbon species onto a
    hydrogen-free diamond (110) surface.  Subsequent C$_2$ molecules are
    deposited on an initially clean surface, in the vicinity of a
    growing adsorbate cluster, and finally, near vacancies just before
    completion of a full new monolayer.  The preferred growth stages
    arise from C$_{2n}$ clusters in near ideal lattice positions forming
    zigzag chains running along the $[\bar{1}10]$ direction parallel to
    the surface.  The adsorption energies are consistently exothermic by
    8--10~eV per C$_2$, depending on the size of the cluster.  The
    deposition barriers for these processes are in the range of
    0.0--0.6~eV.  For deposition sites above C$_{2n}$ clusters the
    adsorption energies are smaller by 3~eV, but diffusion to more
    stable positions is feasible.  We also perform simulations of the
    diffusion of C$_2$ molecules on the surface in the vicinity of
    existing adsorbate clusters using an augmented Lagrangian penalty
    method.  We find migration barriers in excess of 3~eV on the clean
    surface, and 0.6--1.0~eV on top of graphene-like adsorbates.  The
    barrier heights and pathways indicate that the growth from gaseous
    dicarbons proceeds either by direct adsorption onto clean sites or
    after migration on top of the existing C$_{2n}$ chains.
\end{abstract}

%% 1999 PACS:
% 61.43.Bn Structural modeling: serial-addition models, computer simulation
% 81.05.Tp Fullerenes and related materials; diamonds, graphite
% 81.15.Aa Theory and models of film growth

\section{Introduction}

The growth of ultra-nanocrystalline diamond films (UNCD) from C$_2$
precursors produced by C$_{60}$ fragmentation in hydrogen-poor
plasmas\cite{gruen94a} has recently attracted attention because
of high growth rates and resulting good mechanical and electronic
properties of the films.

There is evidence that the growth proceeds
mainly on the (110) face.\cite{chu92}
Previous studies\cite{horner95,redfern96} explored growth by
dicarbon on the hydrogen-terminated (110) face without hydrogen
abstraction by way of insertion of C$_2$ into C--H bonds on the
surface.  In the present work, we consider deposition steps onto the
clean diamond (110) surface without hydrogen participation.
Starting out from a clean surface we investigate the local atomic
configuration arising from the adsorption of a C$_2$ molecule.
Subsequently, more C$_2$ molecules are deposited in the vicinity
of a previous adsorbate cluster.  By comparing the total energy
of these structures, we identify preferred growth stages as those
arising from C$_{2n}$ clusters in the form of zigzag chains running
along the $[\bar{1}10]$ direction parallel to the surface.

We also perform simulations of the diffusion of C$_2$ molecules
on the surface in the vicinity of existing adsorbate clusters
using a constraint minimization
technique.\cite{ciccotti82,ryckaert85,thijssen99}
The barrier heights and pathways indicate that the growth from
gaseous dicarbons proceeds either by direct adsorption onto clean
sites or after migration above existing C$_{2n}$ chains.

The computational method is briefly outlined in Sec.\ II. 
In Sec.\ III, the properties of clean diamond (110) surfaces are calculated 
using density-functional based tight-binding method and compared
with more elaborate {\it ab initio} results.
The initial adsorption steps for diamond growth are 
studied in Sec.\ IV, followed by an evaluation of
diffusion barriers for C$_2$ on diamond (110) in Sec.\ V. 
We then analyze some molecular dynamics trajectories in
Sec.\ VI.
Finally, we present conclusions in \mbox{Sec.\ VII.}

\section{Computational method}

In performing our atomistic simulations,
we apply the density-functional based tight-binding method
\mbox{(DFTB)}.\cite{pore-constr95}
In order to correctly take into account effects of surface polarization
we have used the self-consistent charge extension of the 
method.\cite{elstner98}
It has been applied successfully to
diamond systems including studies of surfaces and
diffusion problems.\cite{Frauenheim:1998a,Kaukonen:1998}
It reproduces total energy differences between various carbon bulk structures
within 50--100~meV/atom compared to {\em ab initio} calculations
(Ref.~\onlinecite{pore-constr95,Koehler:1995a}).
For surface related effects, the errors are determined to be 
less than 0.2~eV/atom.

The surfaces are simulated using a 2-dimensional slab-geometry with
varying thickness.  The bottom layer is saturated by a fixed monolayer
of pseudo-hydrogen atoms which do not mutually interact.  The lateral
extent of the cell varies from $3 \times 3$ unit cells up to $8 \times
3$ unit cells, where the bigger extent is along the chain direction.
This direction is assigned as $x$-axis.
For all atomic structure calculations, we used the $\Gamma$-point
approximation to sample the Brillouin zone, which amounts to a
$\bbox{k}$-point sampling with as many points as real-space unit cells.
We let the atoms relax in a conjugate gradient scheme.

The diffusion and adsorption runs are done using a constrained
conjugate gradient method. The forces during the conjugate gradient
minimization are modified by method described by Ciccotti et al.\ and
Ryckaert.\cite{ciccotti82,ryckaert85,thijssen99}  The center of mass
of the C$_2$ molecule is moved to a given direction with steps of
0.1~\AA{}.  Because the constraint is to the center of mass movement,
the C$_2$ may rotate or dissociate freely.

\section{Clean diamond (110) surface}

The clean (110) surface of diamond as obtained from bulk cleavage
has one dangling bond per atom pointing at an
angle of 19.5$^\circ$ away from the surface normal.  The atoms are
arranged in zigzag chains flat on the surface directed along the
$[\bar{1}10]$ direction.

For initial studies
we used six bulk monolayers of carbon.
We find that only negligible relaxations take place in the lower two layers
and therefore we use only four carbon monolayers for the remaining
calculations.  The applicability of this size of the super-cell
is tested by comparing the minimum energy geometries of a 
C$_4$ adsorbate on the (110) surface with super-cells consisting of
four and six monolayers, respectively.
The comparison between fully relaxed structures calculated with both
number of monolayers
yields very small differences in the atomic positions of about 0.015~\AA{}
in both the lateral and vertical positions of all corresponding atoms,
including those of the adsorbate.

There has been some controversy in the literature about whether the
C(110) surface reconstructs or not.\cite{davidson94,alfonso95}
We find here in agreement with {\em ab initio} studies\cite{kern97} and
experiments\cite{lim-stallcup98} a symmetric flat $(1\times 1)$ surface.
The upper surface layer moves inward by $\Delta z_1 = -0.16$~\AA,
and the next layer outward by $\Delta z_2 = 0.02$~\AA.  The chains
in the top layer are slightly straightened by a relaxation of 0.1~\AA{}
towards the chain axis, resulting in a bond angle of 122.0$^\circ$
and a bond length of 1.44~\AA.
This geometry is in excellent agreement with the {\em ab initio}
results in Ref.~\onlinecite{kern97}, summarized in Table~\ref{tab-geom}.
Taking into account the 0.015~{\AA} offset in the equilibrium
bulk bond length $d_0$, all bond lengths agree within 0.01~\AA,
or 0.6\% of $d_0$.

\section{Adsorption and energetics of small carbon clusters on (110) diamond}

The energetically most favorable cluster configurations after repeated
C$_2$ additions are summarized in Fig.~\ref{fig-overview} up to C$_8$
on (110).  In the following, a more detailed description of the 
surface cluster geometries will be given.

\subsection{Initial C$_2$ deposition}

In order to sample the energy landscape above the clean (110) diamond
surface for C$_2$ adsorption we place a C$_2$ molecule in vertical
orientation near the surface on a hexagonal-like set of points above the
atoms and bond centers of the two topmost monolayers.
By symmetry, only seven positions
remain unique.  The lower atom of the molecule is placed
about 2~\AA{} away from the nearest surface atom.
A conjugate gradient relaxation from each of the lateral starting
positions shows that the molecule is either reflected from
or adsorbed onto the surface.  The reflections occur for starting
positions directly above the atoms and bonds of the top monolayer.
From all of the starting positions not directly above a top-layer chain the
C$_2$ molecule is bonded and forms a bridge above the ``trough''
between two adjacent top-layer chains.  The deposition proceeds
in two stages, see Fig.~\ref{fig-initial}.

Initially, the C$_2$ sticks with one end to the
nearest surface atom with an inclination of about 45$^\circ$ to the
surface normal.  There is a very low energy barrier
of order 0.1~eV towards the
final adsorption stage in which the molecule bonds symmetrically in
an orientation corresponding to the diamond-lattice.  At this final stage,
both adsorbate atoms are 1.0~\AA{} above the top monolayer and
are 3-fold coordinated with a bond length of 1.38~\AA{} between them.
Each adsorbate atom forms 2 bonds towards
the surface:  one bond corresponding to the diamond lattice with a
length of 1.49~\AA, the other being a stretched bond of 1.91~\AA{}
towards the adjacent atom in the same surface chain, see
Fig.~\ref{fig-overview} (a).
The result is the formation of
two 5-fold rings with a common bond formed by the adsorbed molecule.
There is a slight lateral
pinch contraction of the surface monolayer.  All its atoms remain
bonded to the subsurface, which is indented below the
adsorbate by about 0.05~\AA.

The binding energies of the C$_2$ molecule to the clean (110)
surface are fairly high, as can be seen in the fifth column of
Tab.~\ref{tab-ads}.  The energy gain is mainly the result of
forming four bonds to the surface which yields about 2~eV for each bond,
which is plausible by its
similarity to the atomic binding energy of 2.3~eV/bond as obtained
by the DFTB method for bulk diamond.
Breaking the initial triple bond within the C$_2$ molecule and the
stretch of surface bonds near the adsorbate offsets the result to the
adsorption energy of 8.1~eV, listed in the table.

\subsection{Addition of C$_2$}

As a next step, we have studied the effect of adding another C$_2$ molecule
near an existing C$_2$ adsorbate.  Obviously, the deposition onto a
site more than one surface lattice spacing away from the
initial adsorbate results in two isolated clusters of similar configuration
as discussed above, unless a topological mismatch prevents the
completion of the ring formation for a near site, as shown in
Fig.~\ref{fig-c2-ngb} (a).  In this case, the strain field introduced
by the second adsorbate results in a bond switch for one of the
backbonds, with an accordingly high adsorption energy of $-7.2$~eV.
For the deposition onto a site of the neighboring chain, the geometry
and energy is essentially the same as for the isolated case, see
Fig.~\ref{fig-c2-ngb} (b).  We note that in this case, the low barriers
found in the initial adsorption process are no longer present, likely
due to the small local strain field induced.  This effect
supports a rapid spread of such C$_2$ adsorption sites across the
surface.

\subsection{C$_4$ clusters}

The highest gain in energy for the second C$_2$ deposition
is obtained at a site directly above
to the first C$_2$ molecule
(see Figs.~\ref{fig-overview} (b) and \ref{fig-c2-top}).
The second molecule bonds
at the neighboring diamond-lattice site along the $[\bar{1}10]$ valley,
in a diamond-like configuration next to the first one and forms a
4 atom long zigzag chain which amounts to a seed for the next monolayer.
At the ends of the new chain, 5-fold rings are formed
similar to the ones in the C$_2$ case.  The ridge of the adsorbate is
a z-shaped symmetric chain of 3 bonds with lengths of 1.40~\AA{} at the ends
and 1.47~\AA{} in the center.  The central two atoms are raised 1.3~\AA{}
above the top monolayer, which corresponds to a slightly outward relaxation
with respect to the ideal lattice sites.  The end atoms are 0.3~\AA{} closer
to the surface.
The center bond is the common side of two adjacent 6-fold rings connecting
the new chain to the surface monolayer.  Topologically, the
rings supporting the C$_4$ adsorbate above the surface form a pyracylene
structure, which is the basic structural element of a C$_{60}$ fullerene.
A local pinch contraction
of the surface monolayer occurs similar to the previous case.
However, the outer bonds of the adsorbate are here with 1.6~\AA{}
more closer in length to actual bonds, at the expense of
the transition to the uncovered parts of the surface, which are now
more stretched to a distance of 1.9--2.0~\AA{}.
Furthermore, we observe the breaking of backbonds
in the middle of the aggregate, resulting in
a cluster of {\sl sp}$^2$-like coordinated atoms arranged in
a dome-like configuration.
The similarity to such a highly stable configuration as in a fullerene
explains the fact
that this C$_4$ cluster represents the highest energy gain for an
approaching C$_2$ molecule among the structures considered in
Tab.~\ref{tab-ads}.

\subsection{C$_6$ and C$_8$ clusters}

\begin{figure}
    \epsfig{file=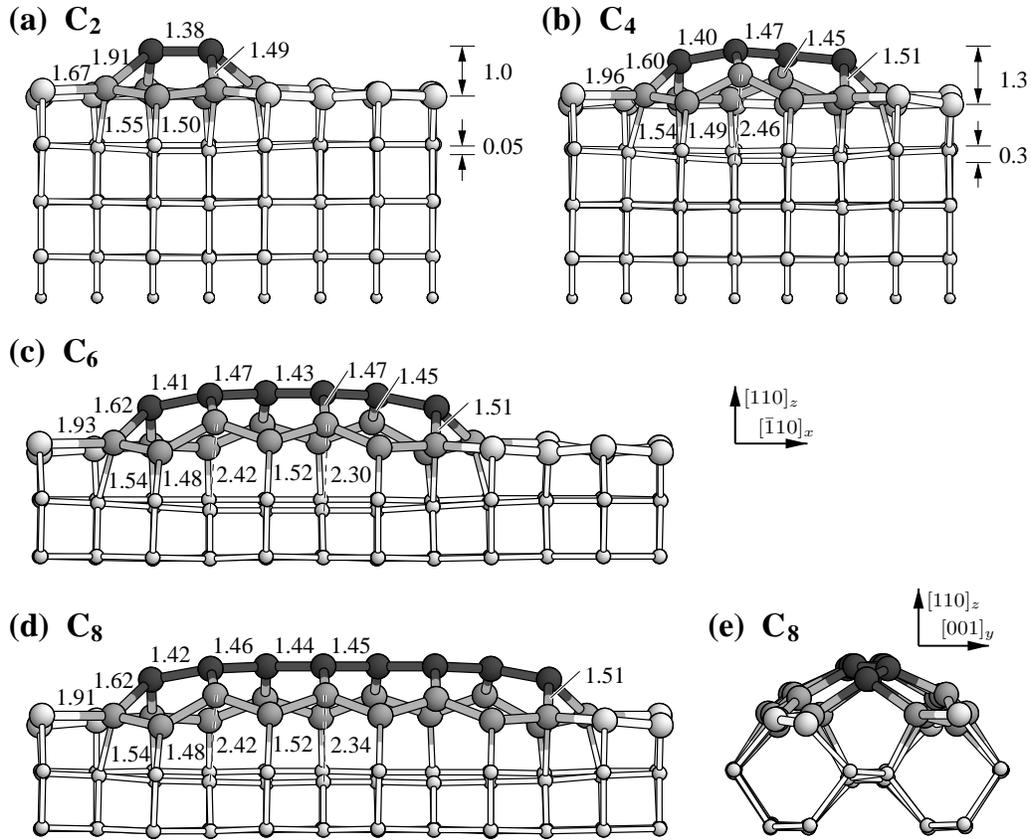, width=12cm, angle=-90}
    \caption{
	Overview of relaxed structures from subsequent depositions of
	C$_2$ molecules onto a clean diamond (110) surface.
	(a) C$_2$, (b) C$_4$, (c) C$_6$, (d) C$_8$, all along
	the $y=[001]$ direction,
	and (e) C$_8$, along the $x=[\bar{1}10]$ direction.
	The numbers given are distances in \AA.
	Atoms of the adsorbate and the surface layer are shown bigger.
	Dark atoms indicate the adsorbate cluster, medium gray atoms
	its first and ring-forming second neighbors within the
	surface layer and lighter gray are other atoms.
	Small spheres at the bottom indicate hydrogen saturation.
	For (c)--(e), only partial models are shown.
    }
    \label{fig-overview}
\end{figure}

The next C$_2$ adsorption to the C$_4$ pyracylene-like adsorbate results
in a C$_6$ adsorbate, shown in Fig.\ref{fig-overview} (c), with an
adsorption energy of 9.6~eV.  Further C$_2$ adsorption yields an
adsorption energy of 8.8~eV and a surface-C$_8$ cluster, shown in
Fig.\ref{fig-overview} (d).  Thus, the energy gain is at least 8~eV
for the repeated C$_2$ surface chain addition, as shown up to C$_8$
on the surface.  We expect the adsorption gain to level off at
about 8.5~eV for longer chains.

The C$_6$ adsorbate has four six-fold rings in the middle and 
two five-fold rings at both ends.  The C$_8$ adsorbate is basically
identical to it, but has of course two more 6-fold rings along its length.
As in the case of the C$_4$ adsorbate, the underlying substrate atoms are
raised above the surface
and flattened to {\sl sp}$^2$-like coordination together with the adsorbed atoms.

\subsection{Adsorption barriers}

The energy barriers for the C$_2$ adsorption
(Tab.~\ref{tab-ads}, fourth column) are either zero or very low.
The highest energy barrier in the case of C$_2$ addition at 
the end of C$_8$ is probably a finite size effect.

The initial C$_2$ addition at a diamond site in a 
``valley'' on the surface makes the bonds shorter in the 
neighboring ``valley''.
This makes the adsorption of an additional C$_2$ easier
to the neighboring `valley' at a diamond site, as can be seen 
by comparing the energies in third and fourth rows 
in Tab.~\ref{tab-ads}.

\subsection{Surface defect formation}

There is a meta-stable energy minimum when adsorbing a
C$_2$ molecule on the top of C$_{4}$ or C$_{6}$ adsorbates
(rows 6 and 8 in Tab.~\ref{tab-ads}).
In the meta-stable state (C$_2$ + C$_{2n}$)
one of the C$_2$ atoms is 
singly bonded to the surface C$_{2n}$-complex, see Fig.~\ref{fig-c8-meta} (b).
The bonded C$_2$ is only 2.0~eV lower in energy than free C$_2$.
We believe this meta-stable minimum configuration plays a key role 
in the growth of (110) diamond.  It enables the diffusion of the 
C$_2$ molecule to the end of an existing C$_{2n}$-complex growing
diamond.
If the C$_2$ molecule is forced deeper onto a C$_{6}$, there is another 
meta-stable minimum energy structure consisting of one 
7-ring, one 6-ring and six 5-rings (row 9 in Tab.~\ref{tab-ads}).
The energy gain from the gaseous C$_2$ to this non-diamond-growth
favoring configuration is 4.9~eV and the energy barrier towards the 
final meta-stable minimum is 1.8~eV.
We believe that similar meta-stable defect structures
form, when a C$_2$ adsorbs to C$_{2n}$ with too high kinetic energy,
approximately $E_K >$ 3--5~eV, taking into account kinetic 
contributions to the energy barriers.

\subsection{Surface vacancy filling}

When the growth proceeds further, different C$_{2n}$ clusters
along the same (110) surface trough will eventually meet and coalesce.
Given that growth proceeds by C$_2$ addition, the critical stage is
reached just before coalescence, when there will be a gap between
two cluster-ends corresponding to either 3 or 2 missing atoms.
Assuming the clusters are seeded at random sites, both cases have
equal probability, but quite different energetics for subsequent
C$_2$ additions.

\begin{figure}
    \epsfig{file=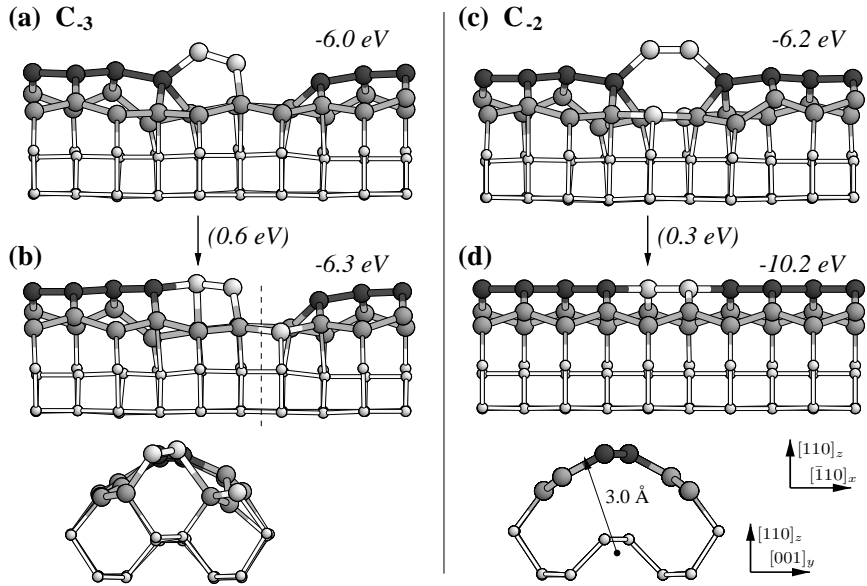, width=12cm}
    \caption{
	Final stages of $[\bar{1}10]$ chain growth and coalescence
	on a diamond (110) surface.
	(a) and (b) for a 3-site vacancy, (c) and (d) for a
	2-site vacancy.  Panels (a) and (c) show the meta-stable
	adsorption phases, and panels (b) and (d) the relaxed minimum
	configuration.
	Atom designation is the same as in Fig.~\ref{fig-overview},
	with the added C$_2$ molecule shown in white.
	The total energies are given relative to initially separated
	substrate and added C$_2$.
	Energies in parentheses indicate barriers.
	The dashed line in (b) indicates a cut used for the
	alternative view along $[\bar{1}10]$ in this panel.
	The arrow in (d) indicates an empirical
	bending radius for the graphene sheet.
    }
    \label{fig-vacfill}
\end{figure}

In either case we see the approaching C$_2$ first in a meta-stable
bridging configuration from a chain end to a bare surface site,
and directly between the two chain ends, for the 3 and 2 site wide
gap, respectively, see Fig.~\ref{fig-vacfill} (a) and (c).
Both added atoms remain just 2-fold coordinated.
After overcoming a barrier of 0.6~eV and 0.3~eV, respectively, the
adsorbate extends the existing C$_{2n}$ chain by another atom pair,
see Fig.~\ref{fig-vacfill} (b) and (d).

For the original 3-atom gap, a single-atom vacancy remains next
to a still just 2-fold coordinated atom.  Accordingly, the gain in
adsorption energy is with 6.3~eV rather low.  However, the remaining
single-atom gap remains reactive, and may be filled at a later stage.

The 2-atom gap yields a much higher adsorption energy of 10.2~eV,
comparable to the high gains found in the initial adsorption stages.
The final stable configuration is a contiguous chain with
broken backbonds for atoms on either side of the top ridge.  This
structure is a bent graphene sheet with a bending radius of about
3~\AA.  Before discussing its properties in the next chapter,
we briefly sketch the other variants for surface vacancy filling.

For the final stage of surface coalescence, we considered a nearly
complete surface monolayer, with up to 3 consecutive atoms along the
$[\bar{1}10]$ direction removed.  The
filling of these surface vacancies results in energy gains between 6.9
and 10.4~eV, as listed in the last rows of Tab.~\ref{tab-ads}.  The
filling of the last single-atom vacancy can take place either by
by a single C atom adsorption to the vacancy, with an energy gain of
10.4~eV without a barrier, or by a C$_2$ addition process, which has a
rather high barrier from a meta-stable minimum at a gain of 6.8~eV to its
completion at 8.4~eV.  Furthermore, it leaves a singly bonded C atom
on otherwise perfect (110) surface.  The energy required to desorb the
extra C atom is of order 8~eV.

\subsection{Graphitisation and rebonding}

\begin{figure}
    \epsfig{file=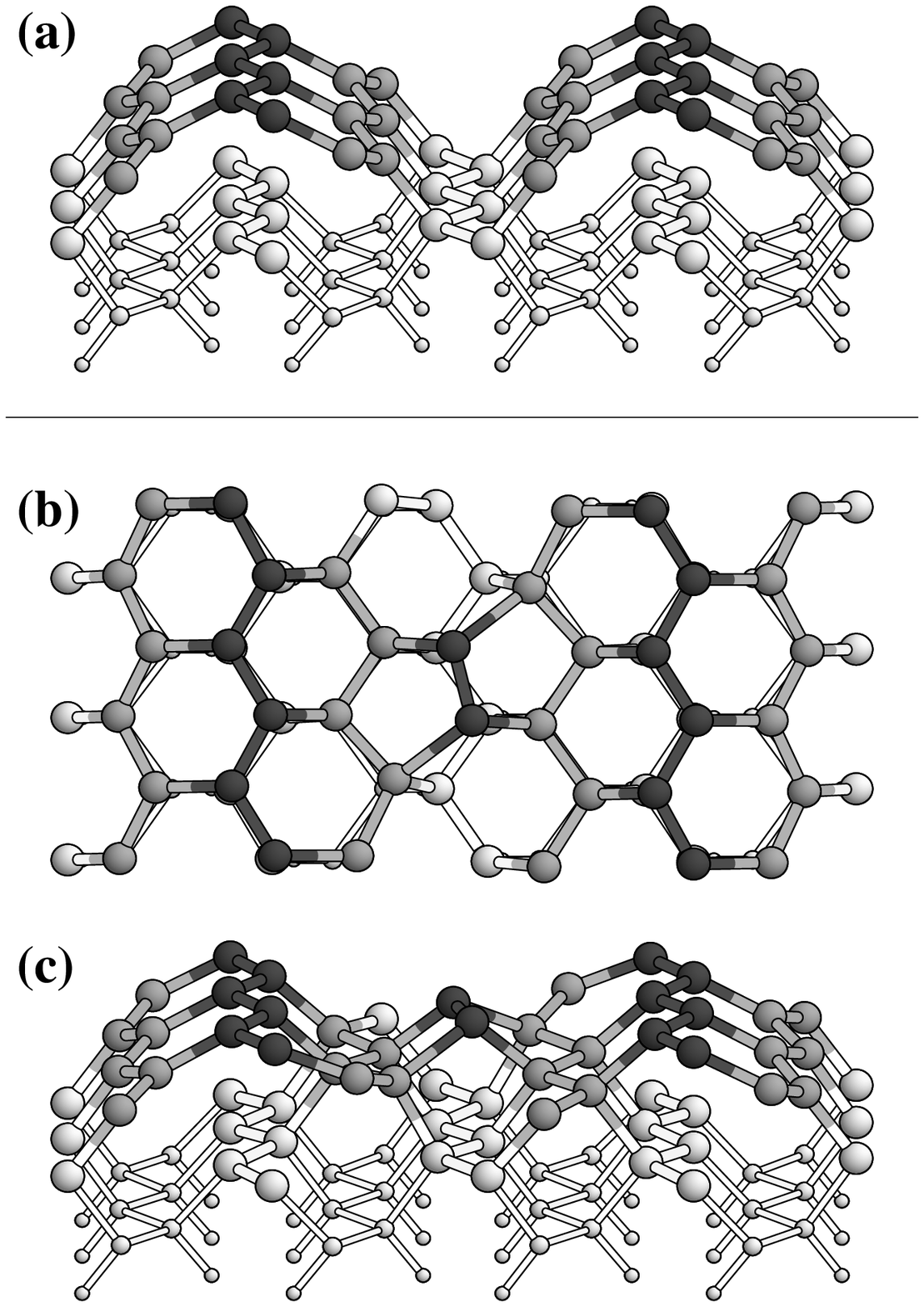, width=8cm}
    \caption{
	Graphitisation on a 50\% covered diamond (110)
	surface in $(2\times 1)$ reconstruction (a),
	and induced rebonding after deposition of C$_2$, (b)
	in top view, (c) in side view.
	Dark, grey and white atoms indicate atoms in the top three monolayers,
	which are also shown in bigger size than the remaining atoms.
	The bottom monolayer is the hydrogen termination.
    }
    \label{fig-rebond}
\end{figure}

As shown in the preceding paragraphs,
the growth process by the C$_2$ chain addition mechanism will eventually
lead to coalescing chains, with broken backbonds on either side.
In order to investigate the consequences of the broken backbonds
for the surface stability during growth
we have relaxed a model in which every other trough along
$[\bar{1}10]$ was covered with a contiguous chain, resulting in a
C(110):$(2\times 1)$ reconstruction, shown
in Fig.~\ref{fig-rebond} (a).  The relaxed structure shows multiple
bent graphene sheets along $[\bar{1}10]$.
The bending orientation is that of a carbon nanotube of the
$(n,n)$ type, known as armchair tube.\cite{saito92}

It is known that graphitisation on clean diamond surfaces, namely on (111)
and near (111) twin boundaries, leads to delamination.\cite{Jungnickel:1995}
We find for the present configuration that it is stable and does not debond.
Furthermore, continued adsorption of C$_2$ in the valley between two
arches is possible without a barrier and more importantly, it
causes the {\sl sp}$^2$-like atoms near the adsorbate to return to an
{\sl sp}$^3$ configuration and rebond in the diamond structure, as is
illustrated in Fig.~\ref{fig-rebond} (b).

\section{Surface diffusion of C$_2$}

In order to estimate the influence of surface diffusion of C$_2$ on
the growth mechanism, we investigated some diffusion paths, as
summarized in Tab.~\ref{tab-dif}.
The associated diffusion barriers along the various C$_2$-related
diffusion paths are shown in the second column of this table, and
the gain in energy in the last column.

Generally, on the clean diamond (110) surface,
the diffusion barriers are rather high, and exceed 3~eV
(Tab.~\ref{tab-dif}, rows 1--3).
This is easily understood by  the strong covalent bond that is
formed between a C$_2$ ad-species and the surface
once the molecule reaches the surface.
The only exception to such high barriers are for sites above
existing adsorbates where the binding energy for further ad-species
is low to begin with.  C$_2$ has the lowest diffusion
barriers when it starts diffusion on top of an existing C$_{2n}$
complex.  In this case, the barrier for diffusion along the adsorbate
ridge is of the order of 1~eV.  The C$_2$ remains nearly vertically
oriented, with one of its atoms bonded to one or two surface atoms
throughout the diffusion path.  The energy barriers
are decreasing when the chain end is approached.
The last energy barrier towards completing a chain addition
step is only 0.6~eV.

We note that the energy gain attainable for a C$_2$ molecule
by diffusion to the end of an existing C$_{2n}$ adsorbate is
considerable, see Tab.~\ref{tab-dif}, last column.  These gains, when
added to the adsorption energies found for the ``top'' deposition
sites, as listed in Tab.~\ref{tab-ads}, naturally result in the same
total adsorption energies as those for the ``end'' sites.  We have
thus found two different growth channels, converging to the same
growth mechanism.  One is adsorption-dominated growth on nearly
clean surfaces with deposition directly into diamond lattice sites,
and the other diffusion-driven on surfaces covered densely with
adsorption clusters.

In the diffusion studies, we found that when the C$_2$ is directed in
both major diffusion directions, either along the chain
direction of C$_{2n}$, or perpendicular to it,
it passes metastable minima before reaching the lowest energy
configurations at the chain ends.
Along the parallel diffusion path, the last metastable
minimum consists of a C$_2$ fragment singly bonded to the
surface, similar to the configuration shown in Fig.~\ref{fig-c2-top}~(b).
For the perpendicular direction away from the chain axis towards
a neighboring surface valley the meta-stable minimum is such that
the C$_2$ atoms complete a 5-fold ring, with each one being
singly bonded to the surface.

While both intermediate and final diffusion barriers are incidentally
the same for diffusion parallel and orthogonal to the C$_{2n}$
chain direction, the ultimate energy {\em gain} is higher by 0.6~eV for
the growth-favoring parallel diffusion; this indicates high adsorption
rates in either case, with a preference towards cystalline growth.
However, the introduction of defects is quite easily possible under this
regime, which helps to explain the rather small grain size found in the
final material.

\section{Molecular dynamics depositions}

Inspired by the adsorption and diffusion results discussed above
we simulated the deposition of C$_2$ on top of a C$_6$-complex on the 
surface using molecular dynamics directly,
though within a rather short time span of just 0.1--0.5~ps.
The goal of these runs was to investigate a diamond growth reaction
as follows:
\begin {equation}
C_{2}^{\rm gas} + C_{2n}^{\rm surf} \rightarrow C_{2n+2}^{\rm surf}.
\end {equation}
We chose the initial kinetic energy for the C$_2$ molecules to
lie in the range of 2--9~eV.

In a first set of experiments,
each molecule is initially aligned orthogonal to
the surface and shot along the direction of the $[\bar{1}10]$ chains
at an angle of 80$^{\circ}$ to the surface normal vector. The
motivation of this particular choice was that the molecule may get
adsorbed at the lowest energy position at the end of a C$_{2n}$
cluster, as identified in the diffusion studies.
All the atoms, except the terminating hydrogen atoms on the
bottom of the surface-slab were allowed to follow the Newtonian
equations of motion.  The molecules with up to 7~eV kinetic energy
resulted in a C$_2$ fragment on top of a C$_{2n}$ cluster, similar
to the structure in Fig.~\ref{fig-c2-top}~(b).  At both
7~eV and 9~eV kinetic energy the molecule was deflected from the C$_{2n}$
cluster but was subsequently adsorbed as a lone C$_2$ on a neighboring
clean surface site.  While the deflection at higher impact energy seems
counter-intuitive at first sight, it must be recalled that the incident
angle is high, so that the nature of the process is rather one of
a steady dissipation of kinetic energy from the approaching C$_2$ into
the substrate until the molecule is slowed down enough to be deposited.

In a second set of runs, a C$_2$ was aligned parallel to the surface
with an initial kinetic energy of 0.1--2~eV and a starting position
on top of a C$_6$ cluster.  This leads again to the C$_2$ fragment
singly bonded to the C$_6$ surface cluster, a configuration from which
diffusion to either end is possible, as established before.
There is a small region above the 
edge of a C$_{2n}$ (n=2,3) cluster, from where a C$_2$, if given
an initial velocity towards to the surface, can bond to the
meta-stable minimum which precedes the diamond position, shown in
Fig.~\ref{fig-c8-meta}.  However, in the molecular dynamics simulations
the barrier towards growth completion is too high in our time-scale,
and we obtained solely the meta-stable minimum.

\section{Summary and conclusions}

We have simulated diamond growth steps by subsequent deposition
of C$_2$ molecules onto a diamond (110) surface.  We find
that the initial C$_2$ adsorption onto a clean (110) diamond
surface proceeds with small barriers (0.1--0.2~eV) into the
diamond lattice site.  The growth mechanism and energetics of
this insertion are similar to those on hydrogenated surfaces, as
suggested previously.\cite{redfern96}  Subsequent C$_2$ additions
on and around the initial adsorbate preferably leads to C$_{2n}$
chains forming along the $[\bar{1}10]$ direction on the surface.
The adsorption energies, as listed in Tab.~\ref{tab-ads} are in the
range of 7--10~eV per C$_2$ molecule at adsorption sites which lead to
chain growth, and slightly smaller, 5--7~eV, for sites leading to
defected growth.
Some backbonds at the C$_{2n}$ chains are broken, leading to a graphene-like
morphology, if a 50\% coverage is reached for the added monolayer.
However, the surface remains stable at this point.  If the C$_2$
deposition continues, it induces healing of
the broken backbonds due to re-formation of {\sl sp}$^3$ bonds at the
terminus of the graphene sheets.

We also find some meta-stable C$_2$ defects during the growth, which
may be responsible for starting new nucleation sites, a tendency that
would explain the rather small grain size in the experimental studies
which motivated this work.  Low energy growth may be possible, if the
approaching molecule has a kinetic energy within a window of 3--5~eV
in order to overcome barriers and avoid defect formation.  The direct
adsorption into a diamond lattice position is possible only at the end
of a C$_{2n}$ cluster or onto a clean site of the surface.
Upon coalescence of different C$_{2n}$ chains, the remaining vacancies
can be filled by the same growth species, although with slightly
higher barriers (0.3--0.6~eV) than in the initial stages.

We have also carried out surface diffusion studies.
Because the C$_2$ adsorptions normally result in tightly bonded adsorbate
structures, inter-island diffusion of C$_2$ molecule is rather unlikely.
However, an intra-island diffusion path exists, where an added C$_2$ molecule
diffuses on top of a C$_{2n}$ chain until it reaches its end and is
incorporated there.  This diffusion behavior supports the C$_2$ addition
model which was evident from the deposition energetics.
An implication of this fact is that an enhancement of surface diffusion
rates would result in an increase in growth rate, by virtue of diffusion
of migrating C$_2$ species to and eventual incorporation into growth
sites.

\section*{Acknowledgments}

This work was carried out with the support of Helsinki University of
Technology, The Center for Scientific Computing in Finland and the
Deutsche Forschungsgemeinschaft.

% TABLES
\newpage
\begin{table}
    \caption{
	Calculated geometry of the clean relaxed C(110) surface.
	Following the notation of Ref.~\protect\onlinecite{kern97},
	$d_{ij}$ is the bond length between atoms of the $i$\/th and $j$\/th
	layers; $d_0$ is the equilibrium bulk bond length, and
	$\theta_i$ is the bond angle within layer $i$.
	$\Delta y_i$ and $\Delta z_i$ are the
	relaxation of the atomic positions in layer $i$ along the y ($[001]$,
	transversal to chains) and z ($[110]$, surface-normal) directions,
	respectively.  The units are {\AA} and degrees, respectively.
    }
    \label{tab-geom}
    \begin{tabular}{lrrrr}
	& \multicolumn{2}{c}{Ref.~\onlinecite{kern97}}
	& \multicolumn{2}{c}{this work} \\
    \tableline
    $d_{0}$	& 1.529	& (100\%)	& 1.544 & (100\%) \\
    $d_{11}$	& 1.419	& ($-7.2\%$)	& 1.441 & ($-6.7\%$) \\
    $d_{12}$	& 1.467	& ($-4.1\%$)	& 1.490 & ($-3.5\%$) \\
    $d_{22}$	& 1.490	& ($-2.6\%$)	& 1.505 & ($-2.6\%$) \\
    $d_{23}$	& 1.576	& ($+3.1\%$)	& 1.587 & ($+2.7\%$) \\
    $d_{33}$	& 1.526	& ($-0.2\%$)	& 1.542 & ($-0.2\%$) \\
    \tableline
    $\theta_{1}$	& \multicolumn{2}{d}{123.3}
    			  & \multicolumn{2}{d}{122.0}	\\
    $\theta_{2}$	& \multicolumn{2}{d}{113.8}
    			  & \multicolumn{2}{d}{113.8}	\\
    $\Delta y_{1}$	& \multicolumn{2}{c}{$\pm 0.10$}
    			  & \multicolumn{2}{c}{$\pm 0.10$}\\
    $\Delta z_{1}$	& \multicolumn{2}{c}{$-0.17$}
    			  & \multicolumn{2}{c}{$-0.16$}	\\
    $\Delta y_{2}$	& \multicolumn{2}{c}{$\pm 0.03$}
    			  & \multicolumn{2}{c}{$\pm 0.00$}\\
    $\Delta z_{2}$	& \multicolumn{2}{c}{$+0.03$}
    			  & \multicolumn{2}{c}{$+0.02$}	\\
    \end{tabular}
\end{table}

\begin{table}
\caption{
    Energy barriers (E$_{\rm barr}$) and adsorption energies (E$_{\rm ads}$)
    of C$_2$ (and C) at the initial and final stages of growth, for varying
    target sites.  The ``top'' position means that the initial
    position of the C$_2$ is above a C$_{2n}$ cluster; ``end''
    refers to an initial C$_2$ position above the edge along the
    $[\bar{1}10]$ direction of the C$_{2n}$ cluster.  ``same'' and
    ``other'' express whether the added  C$_2$ is above the same or the
    adjacent (110) trough as the existing C$_{2n}$ cluster on the
    surface.  Negative indices indicate missing C atoms in an otherwise
    contiguous chain or on the (110) surface.
}
\label{tab-ads}
\begin{tabular}{lllccc}
Initial Configuration
    & Final Configuration	& Figure
	& E$_{\rm barr}$/eV	& E$_{\rm ads}$/eV	& E$_{\rm ads}$/eV \\
	    &&&&& (Ref.~\protect\onlinecite{redfern96}) \\
\tableline
 C$_2$ + (110)
    & (110):C$_2$	& {Figs.~\ref{fig-overview} (a), \ref{fig-initial}}
 	& 0.1	& $-8.1$\tablenotemark[1] & $-7.8$ \\
 C$_2$ + (110):C$_2$ top
    & (110):C$_4$	& {Figs.~\ref{fig-overview} (b), \ref{fig-c2-top}}
 	& 0.1	& $-10.3$\tablenotemark[1] & $-8.8$ \\
 C$_2$ + (110):C$_2$ same
    & (110):C$_2$,C$_2$ same& {Fig.~\ref{fig-c2-ngb} (a)}
 	& 0.2	& $-7.2$ & \\
 C$_2$ + (110):C$_2$ other
    & (110):C$_2$,C$_2$ other& {Fig.~\ref{fig-c2-ngb} (b)}
 	& 0.0	& $-8.3$ & $-7.8$ \\
\tableline
 C$_2$ + (110):C$_4$ end
    & (110):C$_6$	& {Fig.~\ref{fig-overview} (c)}
 	& 0.1	& $-9.6$\tablenotemark[1] & \\
 C$_2$ + (110):C$_4$ top
    & (110):C$_4$,C$_2$	& --
 	& 0.7	& $-6.5$ & \\
\tableline
 C$_2$ + (110):C$_6$ end
    & (110):C$_8$	& {Fig.~\ref{fig-overview} (d)}
 	& 0.5	& $-8.8$\tablenotemark[1] & \\
 C$_2$ + (110):C$_6$ top
    & (110):C$_6$,C$_2$	& {Fig.~\ref{fig-c8-meta} (b)}
 	& 0.0	& $-2.0$ & \\
 C$_2$ + (110):C$_6$ top
    & (110):C$_8$ defect& {Fig.~\ref{fig-c8-meta} (c)}
 	& 1.8	& $-4.9$ & \\
 C$_2$ + (110):C$_6$ other
    & (110):C$_6$,C$_2$ other& --
 	& 0.0	& $-7.8$ & \\
\tableline
 C$_2$ + (110):$(2\times 1)$:C$_{-3}$
    & (110):$(2\times 1)$:C$_{-1}$ & {Fig.~\ref{fig-vacfill} (b)}
 	& 0.6	& $-6.3$ & \\
 C$_2$ + (110):$(2\times 1)$:C$_{-2}$
    & (110):$(2\times 1)$	& {Fig.~\ref{fig-vacfill} (d)}
 	& 0.3	& $-10.2$\tablenotemark[1] & \\
 C$_2$ + (110):$(2\times 1)$
    & (110):$(2\times 1)$:C$_2$& {Fig.~\ref{fig-rebond}}
 	& 0.0	& $-7.2$ & \\
\tableline

 C$_2$ + (110):C$_{-3}$
    & (110):C$_{-1}$	& --
 	&  0.5 & $-6.9$ & \\
 C$_2$ + (110):C$_{-2}$
    & (110)			& --
 	&  0.4 & $-8.1$ & \\
 C$_2$ + (110):C$_{-1}$
    & (110):C		& --
 	& 0.0\tablenotemark[2]/2.6 & $-6.8$\tablenotemark[2]/$-8.3$ & \\
 C$_{\,\;}$ + (110):C$_{-1}$
    & (110)			& --
 	& 0.0	& $-10.4$ & \\
\end{tabular}
\tablenotetext[1]{C$_{2n}$ chain growth and coalescence processes}
\tablenotetext[2]{meta-stable state}
\end{table}

\begin{table}
    \caption{
	The diffusion barriers E$_{\rm barr}$ and the change in the 
	total energy along the diffusion path.
	The low energy barriers (rows 4 and 5) are associated with the
	diffusion of a vertically aligned C$_2$ on top of a C$_{2n}$ 
	cluster.
    }
    \label{tab-dif}
    \begin{tabular}{lcc}
    Path	& E$_{\rm barr}$/eV & (E$_{\rm final} - $E$_{\rm init}$)/eV \\ 
    \tableline
      C$_2$ along valley
	    & 3.8 & 0.0 \\
      C$_2$ to other valley
	    & 3.3 & 0.0 \\
      C$_2$ to C$_4$ along $\rightarrow$ C$_6$
	    & 3.7 & $-1.8$ \\	
    \tableline
      C$_2$ to end of C$_6$ $\rightarrow$ C$_8$
	    & 1.0\tablenotemark[1]/0.6\tablenotemark[2] & $-3.7$ \\
      C$_2$ to side of C$_6$ $\rightarrow$ C$_8$
	    & 1.0\tablenotemark[1]/0.6\tablenotemark[3] & $-3.1$ \\
    \end{tabular}
    \tablenotetext[1]{near center of C$_6$}
    \tablenotetext[2]{near end of C$_{2n}$}
    \tablenotetext[3]{from meta-stable to final energy minimum}
\end{table}

% FIGURES
\newpage

\begin{figure}
    \epsfig{file=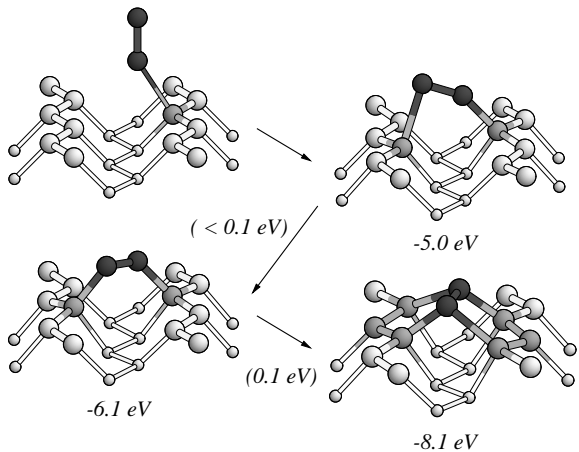, width=8cm}
    \caption{
	Initial steps for deposition of a C$_2$ molecule onto a clean diamond
	(110) surface.  The energies are given relative to a clean surface and
	a distant C$_2$.  Energies in parentheses indicate barriers.
	Atom designation is the same as in Fig.~\ref{fig-overview}.
    }
    \label{fig-initial}
\end{figure}

\begin{figure}
    \epsfig{file=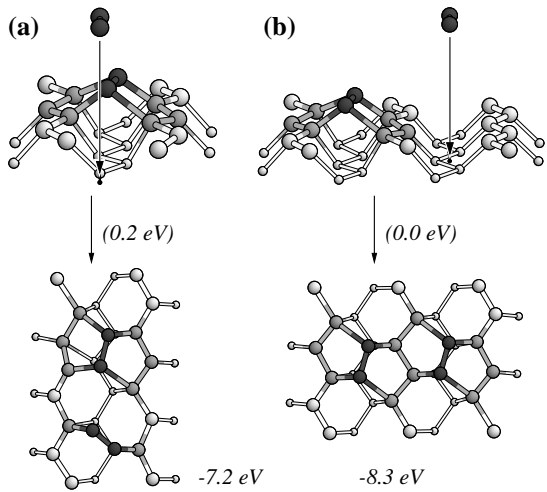, width=8cm}
    \caption{
	Continued deposition of a C$_2$ molecule onto a diamond (110) surface
	on sites {\em next to} an existing C$_2$ adsorbate.
	The targeted neighboring site is
	(a) along the $x=[\bar{1}10]$ direction, and
	(b) along the $y=[001]$ direction.
	The total energies are given relative to initially separated components.
	Energies in brackets indicate barriers.
	Atom designation is the same as in Fig.~\ref{fig-overview}.
	Additional small markers indicate the target location.
    }
    \label{fig-c2-ngb}
\end{figure}

\begin{figure}
    \epsfig{file=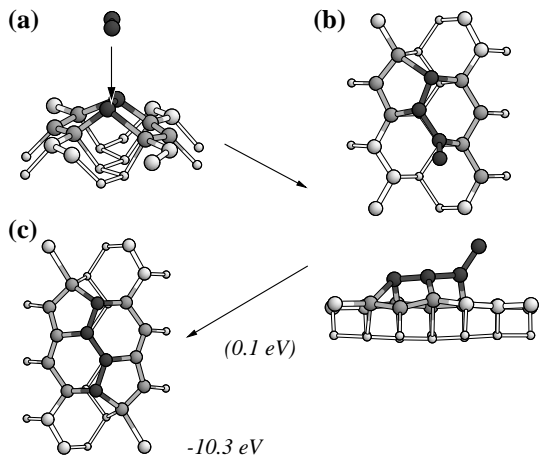, width=8cm}
    \caption{
	Continued deposition of a C$_2$ molecule onto a diamond (110) surface
	on {\em top of} an existing C$_2$ adsorbate, with the transition
	state shown from the top and side.  The resulting
	adsorbate cluster is topologically similar to a C$_{60}$ fragment.
	The panels show
	(a) the initial, (b) the transition, and (c) the final state.
	The total energies are given relative to initially separated components.
	Atom designation is the same as in Fig.~\ref{fig-overview}.
    }
    \label{fig-c2-top}
\end{figure}

\begin{figure}
    \epsfig{file=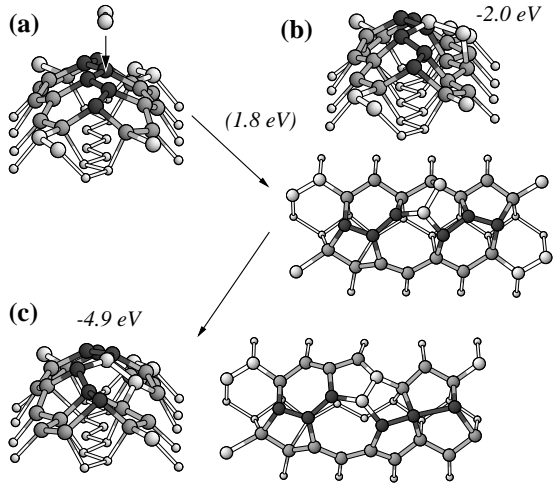, width=8cm}
    \caption{
	Continued deposition of a C$_2$ molecule onto a diamond (110) surface
	over a C$_6$ adsorbate with high insertion energy.  The panels show
	(a) the initial, (b) the transition, and (c) the final state.
	The total energies are given relative to initially separated
	components.
	Atom designation is the same as in Fig.~\ref{fig-overview},
	with the added C$_2$ molecule shown in white.
    }
    \label{fig-c8-meta}
\end{figure}

\begin{references}
\bibitem[*]{mailto}
	Electronic address: sternberg@phys.uni-paderborn.de
\bibitem{gruen94a}
	%D.M. Gruen, S. Liu, A.R. Krauss, J. Luo and X. Pan,
	D.M. Gruen {\em et al.,}
	 {J. Appl. Phys.} {\bf 64}, 1502 (1994).
	% experiments.
\bibitem{chu92}
	C.J. Chu, R.H. Hauge, J.L. Margrave and M.P. D'Evelyn,
	{Appl. Phys. Lett.} {\bf 61} (12), 1393 (1992).
	% experiments.
\bibitem{horner95}
	D.A. Horner, L.A. Curtiss and D.M. Gruen,
	{Chem. Phys. Lett.} {\bf 233}, 243 (1995).
	% semiemp. clusters.
\bibitem{redfern96}
	P.C. Redfern, D.A. Horner, L.A. Curtiss and D.M. Gruen,
	{J. Phys. Chem.} {\bf 100}, 11654 (1996).
	% C2 insertion onto 110:H
\bibitem{ciccotti82}
        G. Ciccotti, M. Ferrario and, J.-P. Ryckaert,
        {Mol. Phys.} {\bf 47}, 1253 (1982).
        % constraint force 1
\bibitem{ryckaert85}
         J.-P. Ryckaert,
        {Mol. Phys.} {\bf 55}, 549 (1985).
        % constraint force 2
\bibitem{thijssen99}
        J.M. Thjissen,
        {\em Computational Physics}
        (University Press, Cambridge, 1999) pp. 218--220.
        %  constraint force 3
\bibitem{pore-constr95}
	%D. Porezag, Th. Frauenheim, Th. K\"ohler, G. Seifert and R. Kaschner,
	D. Porezag {\em et al.,}
	{Phys. Rev. B\/} {\bf 51}, 12947 (1995).
	% scf contruction of LCAO basis
\bibitem{elstner98}
	M. Elstner {\em et al.,}
	{Phys. Rev. B} {\bf 58}, 7260 (1998).
	% scc-DFTB
\bibitem{Frauenheim:1998a}
	Th. Frauenheim {\em et al.,}
	{Diam. \& Rel. Mat.} {\bf 7}, 348 (1998).
	% surf. and diffusion
\bibitem{Kaukonen:1998}
	M. Kaukonen {\em et al.,}
	{Phys. Rev. B} {\bf 57}, 9965 (1998).
	% constrained C.G.
\bibitem{Koehler:1995a}
	Th. K{\"o}hler, Th. Frauenheim and G. Jungnickel,
	{Phys. Rev. B} {\bf 52}, 11837 (1995).
	% DFTB general method
\bibitem{davidson94}
	B.N. Davidson and W.E. Pickett,
	{Phys. Rev. B} {\bf 49}, 11253 (1994).
	% 110 structure, TB
\bibitem{alfonso95}
	D.R. Alfonso, D.A. Drabold and S.E. Ulloa,
	{Phys. Rev. B} {\bf 51}, 14669 (1995).
	% 110 structure, TB
\bibitem{kern97}
	G. Kern and J. Hafner,
	{Phys. Rev. B} {\bf 56}, 4203 (1998).
	% 110 structure, VASP
\bibitem{lim-stallcup98}
	S.C. Lim, R.E. II Stallcup, I. Akwani and J.M. Perez,
	in {\em Materials Issues in Vacuum Microelectronics,}
	Symp. Mater. Res. Soc, 165 (1998).
	% 110 expt.
\bibitem{saito92}
	R. Saito, M. Fujita, G. Dresselhaus and M.S. Dresselhaus,
	{Appl. Phys. Lett.} {\bf 60}, 2204 (1992).
	% carbon nanotube conductivity
\bibitem{Jungnickel:1995}
	%G. Jungnickel, D. Porezag and T. Frauenheim,
	%W. R. L. Lambrecht, B. Segall, and J. C. Angus,
	G. Jungnickel {\em et al.,}
	{MRS Symp. Proc.} {\bf 383}, 349 (1995).
	% (111) delamination
\end{references}
\end{document}